\UseRawInputEncoding
\documentclass[
 reprint,
 amsmath,amssymb,
 aps,
 prl,
]{revtex4-2}

\usepackage[utf8]{inputenc}
\usepackage{float}
\usepackage[normalem]{ulem}
\usepackage{graphicx}% Include figure files
\usepackage{dcolumn}% Align table columns on decimal point
\usepackage{bm}% bold math
\usepackage{xcolor}
\usepackage{hyperref}% add hypertext capabilities
\usepackage{subfiles} % Best loaded last in the preamble
\usepackage{newunicodechar}
\newunicodechar{−}{-}
\newcommand{\hs}[1]{\textcolor{black}{#1}}

\begin{document}

\preprint{APS/123-QED}

%%-----------------TITLE + AUTHORS---------------%%
\title{Spontaneous emergence of run-and-tumble-like dynamics in a robotic analog of \textit{Chlamydomonas}: experiment and theory}% 

\author{Somnath Paramanick$^{1}$}
\thanks{Equal contribution}
\author{Umashankar Pardhi$^{2}$}
\thanks{Equal contribution}
\author{Harsh Soni$^{2}$}
\email{harsh@iitmandi.ac.in}
\author{Nitin Kumar$^{1}$}
\email{nkumar@iitb.ac.in}

\affiliation{
	$^1$Department of Physics, Indian Institute of Technology Bombay Powai, Mumbai 400076, India \\
	$^2$School of Physical Sciences, Indian Institute of Technology Mandi, Mandi 175001, India \\
	}
\date{\today}

%-----------------ABSTRACT---------------%%
\begin{abstract}
Run-and-tumble (RT) motion is commonly observed in flagellated microswimmers, arising from synchronous and asynchronous flagellar beating. One such example is a biflagellated alga, called \textit{Chlamydomonas reinhardtii}. Its flagellar synchronization is not only affected by hydrodynamic interactions but also through contractile stress fibers that mechanically couple the flagella, enabling adaptable swimming behaviour. To explore this, we design a macroscopic mechanical system that comprises dry, self-propelled robots linked by a rigid rod to model this organism. By varying the attachment points of the two ends of the rod on each robot, the model incorporates the effect of fiber contractility observed in the real organism. To mimic a low Reynolds number environment, we program each robot to undergo overdamped active Brownian (AB) motion. We find that such a system exhibits RT-like behavior, characterized by sharp, direction-reversing tumbles and exponentially distributed run times, consistent with the real organism. Moreover, we quantify tumbling frequency and demonstrate its tunability across experimental parameters. Additionally, we provide a theoretical model that reproduces our results, elucidating physical mechanisms governing RT dynamics. Thus, our robotic system not only replicates RT motion but also captures several subtle characteristics, offering valuable insights into the underlying physics of microswimmer motility.

%Drawing inspiration from the motility behaviour of microorganisms, we introduce a highly tunable, robotic system self-actuating into the run-and-tumble (RT)-like motion. It comprises two disk-shaped, centimeter-scale programmable robots individually programmed to perform overdamped active Brownian (AB) motion and connected by a rigid rod. The rod is attached to pivot points located on off-centered, mirror-symmetric points on each robot, allowing for its free rotation at both ends. We show that the collective dynamics of this system execute RT-like motion with characteristic sharp tumble events and exponentially distributed run times, similar to those observed in microorganisms. We further quantify emerging dynamics in terms of tumbling frequency and tune it over a wide range of experimental parameters. We also develop a theoretical model that reproduces our experimental results and elucidates the underlying physical mechanisms governing the rich phase behavior of RT motion.
\end{abstract}

\maketitle

%\tableofcontents

%% INTRODUCTION

Motility is one of the defining features of active and living organisms across length scales \cite{ramaswamy2010mechanics, gompper2021motile, berg2000motile}. One prominent example is run-and-tumble (RT) motion commonly seen in swimming microorganisms living in a low Reynolds number environment \cite{berg1972chemotaxis, polin2009chlamydomonas, li2008persistent, findlay2021high, purcell2014life}. Here, organisms trace relatively straight paths (runs) before abruptly changing to a new, randomly chosen direction (tumbles). It is known to originate from coordinated dynamics of multiple active units shaped as filamentous appendages, called flagella or cilia, decorating the body of microorganisms. These units exhibit intrinsic activity through their rhythmic beating, propelling the organism forward in a series of synchronous and asynchronous cycles \cite{polin2009chlamydomonas, LaugaPhysicsToday2012, goldstein2011emergence, brumley2014flagellar, wan2014lag, goldstein2009noise, woolley2009study, mondal2020internal, friedrich2012flagellar}. Two commonly studied model organisms for understanding the emergence of run-and-tumble (RT) motion are \textit{E. coli} \cite{berg1972chemotaxis} and \textit{Chlamydomonas reinhardtii} \cite{polin2009chlamydomonas}.

Most organisms showing RT motion lack cognitive abilities or centralized control systems like the brain. Therefore, simple coupling rules between its active components must govern their dynamics. 
Since they swim in a fluid medium, attempts have been made to investigate the effects of hydrodynamic coupling between beating flagella on motility ~\cite{BennettGolestanianPRL2013, MunJu2004pre, Vilfan2006prl, Niedermayer2008Chaos, Uchida2011prl, Oliver2018SoftMatter}. However, there is growing evidence that mechanical couplings between active components inside the organism's body also influence the swimming behaviour. This is especially true for biflagellated organisms like \textit{Chlamydomonas} \cite{friedrich2012flagellar, quaranta2015hydrodynamics,wan2016coordinated,quaranta2015hydrodynamics, geyer2013cell,soh2022intracellular}. The basal bodies of their flagella are connected through a contractile fiber, called the distal fiber,
that is known to influence flagellar orientation and affect swimming behaviour \cite{hayashi1998real, mcfadden1987basal}. Therefore, it remains to be seen whether an artificial model system that demonstrates and validates the emergence of adaptable RT motion based on these mechanisms can be envisaged.

%%-----------------FIGURE I---------------%%
\begin{figure}[t]
	\includegraphics[width= 0.482\textwidth]{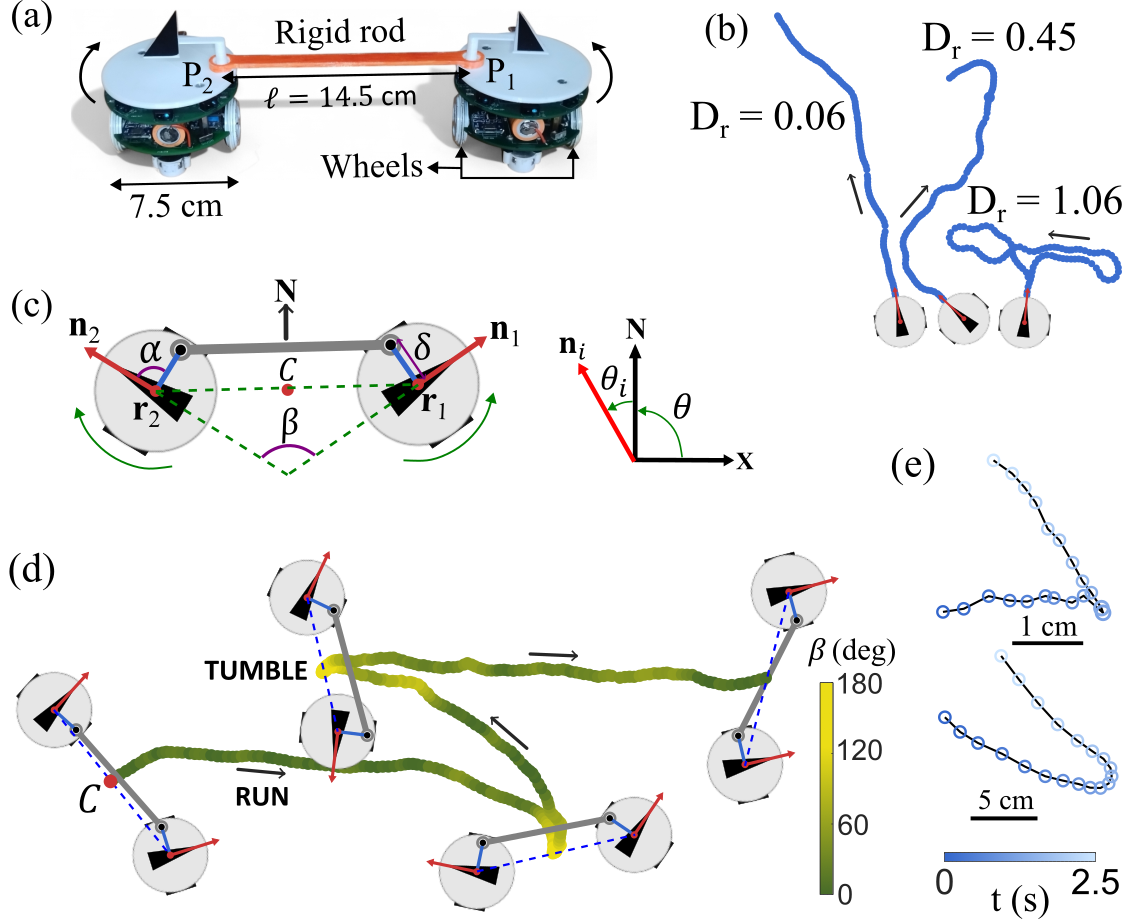}
\caption{\textbf{(a)} Photograph of the experimental system featuring coupled robots. The pivot points, $P_1$ and $P_2$, allow the connecting rod to rotate freely in the horizontal plane. \textbf{(b)} Typical trajectories of free robots executing overdamped AB motion with $v_a$ = 5 cm s$^{-1}$ for three different values of the rotational diffusion constant, $D_r$ (in rad$^2$ s$^{-1}$). \textbf{(c)} A schematic diagram highlighting key variables incorporated into the theoretical model. \textbf{(d)} A typical RT trajectory of the centroid point $C$ in the direction of black arrows, with $v_a$ = 5 cm s$^{-1}$, $D_r$ = 0.06 rad$^2$ s$^{-1}$, $\delta$ = 3 cm, and $\alpha = 90^\circ$. The color bar shows the value of $\beta$. \textbf{(e)} Zoomed-in views of typical tumble events under conditions of high (top) and low (bottom) substrate friction. We increased $v_a$ to 20 cm s$^{-1}$ for the low friction case to better illustrate the smoothening of the tumble event.}
	\label{fig1}
\end{figure}
%-----------------------------------------------------%%

In the past, attempts have been made to mimic various dynamical features of microorganisms in synthetic systems \cite{xia2024biomimetic, xu2024constrained, sanchez2011cilia, tiwari2020periodic, han2021low, DauchotNatPhy2022, DuringPRL2024}. Notwithstanding, an artificial analog system that accurately mimics RT motion with quick, straight run trajectories and slow, sharp tumbles is lacking. While there are reports of RT-like motion in vibrated granular particles \cite{kumar2019trapping} and self-propelling camphor boats \cite{PampaCamphor}, these studies often feature long tumble durations and very short runs. This behavior contrasts with microorganisms, where runs are much longer than tumbles \cite{berg1972chemotaxis, polin2009chlamydomonas, li2008persistent}. This discrepancy likely originates due to finite inertial effects in these macroscale synthetic systems \cite{scholz2018inertial, dutta2024inertial, caprini2021inertial, fersula2024self}, limiting their relevance to microscopic organisms that exclusively operate in inertialess limit. Therefore, for a macroscale artificial model system to accurately reproduce RT motion, it must similarly operate in an overdamped regime. To the best of our knowledge, no experimental system currently exists that fulfills these conditions while providing tunable statistical features of the RT motion. In this study, we introduce an experimental model system supported by a theoretical framework where a highly tunable RT-like motion emerges spontaneously.

Our experimental system consists of two self-propelled robots connected with a rigid rod, as shown in Fig.~\ref{fig1}(a). The setup is conceived to model robots as flagella and the rod as the contractile fiber connecting basal bodies within the \textit{Chlamydomonas}. Each robot is 7.5 cm in diameter and individually programmed to execute in-plane overdamped active Brownian (AB) motion along axes passing through their centers, aligned parallel to their wheels \cite{paramanick2024programming, paramanick2024uncovering} [also see Supplemental Material (SM) Section IA \cite{supp_mat}]. This AB motion is characterized by a constant self-propulsion speed $v_a$ and the rotational diffusion constant $D_r$. Both robots are covered with 3D-printed caps featuring black, triangle-shaped markers, for tracking their in-plane positions and orientations. The AB trajectories of the single, non-connected robot are shown in Fig.~\ref{fig1}(b) for three typical values of $D_r$ while keeping $v_a$ = 5 cm s$^{-1}$ (see SM movie 1). Central to their overdamped feature is the fact that they propel forward using their wheels, rolling without slipping on a flat glass surface covered with a white sheet of paper. This provides sufficient traction for the wheels, making the effects of inertia negligible. The connecting rod of length $\ell = 14.5$ cm is free to rotate about pivot points ($P_1$ and $P_2$) located on the off-centered, mirror-symmetric points on each robot [Figs.~\ref{fig1}(a) $\&$ (c)]. Here, $\delta$ is the distance of these points from robots' centres, and $\alpha$ denotes the angle between the pivot-to-center line and the polarity axis. We vary $D_r$, $\delta$ and $\alpha$ between 0.06 - 1.33 rad$^2$s$^{-1}$, $1$ - $3$ cm and $30^\circ$ - $150^\circ$ respectively. $v_a$ is kept fixed at 5 cm s$^{-1}$. A combination of self-propulsion and constraint force due to the rigid rod generates net torque around vertical axes passing through $P_1$ and $P_2$. These torques act stochastically in opposite handedness for the two robots indicated by curved arrows in Fig.~\ref{fig1}(a). To summarize, this setup incorporates the fact that the run and tumble states are associated with translational and rotational motion, which originate from an interplay between active forces and their corresponding moments, respectively. In what follows, we show that this simple-minded design is capable of showing RT-like motion. 

%% THEORY MODEL

We now develop a theoretical model for our system (for details, see SM sections IIA and IIB \cite{supp_mat}). In our model, all dissipative forces, including frictional and resistive effects on a single robot with translational and rotational velocity $\mathbf{v}$ $\&$ $\bm{\omega}$, are represented by a  dissipative force and torque  $ 
    \mathbf{F}_\text{d} = -\bm{\Gamma} \cdot \mathbf{v} $ and $\bm{\tau}_\text{d} = -\Gamma_\tau \bm{\omega}$, respectively.  
Here, $\bm{\Gamma}$ and $\Gamma_\tau$ denote the translational dissipation tensor and rotational dissipation coefficient, respectively. We assume that even when connected, robots remain overdamped. We also neglect the leading-order dependence of $\bm{\Gamma}$ and $\Gamma_\tau$ on $\mathbf{v}$ and $\bm{\omega}$. In experiments, the friction perpendicular to a robot’s polar axis (from wheel sliding) is significantly greater than the parallel friction (from rolling). Consequently, $\bm{\Gamma}$ adopts a non-scalar form dependent on the robot’s orientation $\mathbf{n}$, with dissipative forces given by $-\Gamma_\parallel \mathbf{v}_\parallel$ and $-\Gamma_\perp \mathbf{v}_\perp$, where $\mathbf{v}_\parallel$ and $\mathbf{v}_\perp$ are the velocity components along and perpendicular to $\mathbf{n}$, and $\Gamma_\parallel, \Gamma_\perp > 0$. 
Using these assumptions, we derive the equations of motion for the orientation $ \theta $ of the unit vector $ \mathbf{N} $, which is normal to the rod, the orientation angles $ \theta_i $ of the robots ($ i = 1,2 $) relative to $ \mathbf{N} $, and the centroid, $C$, of the system, $ \mathbf{r} = (\mathbf{r}_1 + \mathbf{r}_2)/2 $, where $ \mathbf{r}_i $ represents the center's position of $i$th robot [see Fig.~\ref{fig1}(c) and Eqs.~(30–32) of SM \cite{supp_mat}]. All parameters match experimental values, except for two phenomenological ratios adjusted for optimal agreement: $\Gamma_\parallel / \Gamma_\perp = 0.1$ and $4\Gamma_\tau / (\Gamma_\parallel d^2) = 1$, with robot diameter $d = 7.5$ cm.

%% RESULTS
%Now we present experimental and simulation results. We conduct simulations based on our theoretical model and show their results alongside experiments. 
We start with parameters $\alpha = 90 ^\circ$ and $\delta = 3$ cm. When connected robots are set in motion, the point $C$ exhibits dynamics resembling fast, nearly straight trajectories (runs) with occurrences of abrupt, sharp turns and sudden halts (tumbles). A typical trajectory is shown in Fig.~\ref{fig1}(d) and SM movie 2 for $D_r = 0.06$ rad$^2$ s$^{-1}$ and 1.33 rad$^2$ s$^{-1}$ in SM movie 3. We reproduce these dynamics in simulations \hs{by solving equations of $\theta_i$, $\theta$, and $\mathbf{r}_i$ of} our theoretical model (SM movies 2 and 3). By analyzing extended simulation trajectories, we find that long-time dynamics continue to remain diffusive (See SM section IIC and Fig.~S8 \cite{supp_mat}). We highlight that it is essential to minimize inertial time scales in capturing sudden, sharp tumbling events in experiments. This implies minimizing factors of $m/\Gamma_{\parallel}$, $m/\Gamma_{\perp}$, and $I/\Gamma_{\tau}$, where $m$ and $I$ are the mass and the moment of inertia of each robot, respectively (Eqs. 4 and 5 in SM section IIA \cite{supp_mat}). To achieve the limit where these factors are non-negligible, we performed an experiment on a slippery glass surface coated with coconut oil. Interestingly, we find that a majority of tumbling events become significantly smoother (See Fig.~\ref{fig1}(e) and SM movie 4), making them harder to distinguish from run states. In contrast, such smooth turns are rare on a frictional substrate. These findings are in accordance with previous studies in some active matter experiments where inertial effects are known to introduce time delays preventing abrupt changes in direction \cite{scholz2018inertial, dutta2024inertial, caprini2021inertial, patel2023exact}. We further demonstrate that the RT motion does not occur when individual robots are programmed to perform Brownian dynamics (SM movie 5 and SM section IB \cite{supp_mat}), confirming its inherently active nature \cite{maitra2024activity}. Henceforth, all results are obtained from experiments conducted on a frictional surface.

%%-----------------FIGURE II---------------%%
\begin{figure}[t]
	\includegraphics[width= 0.485\textwidth]{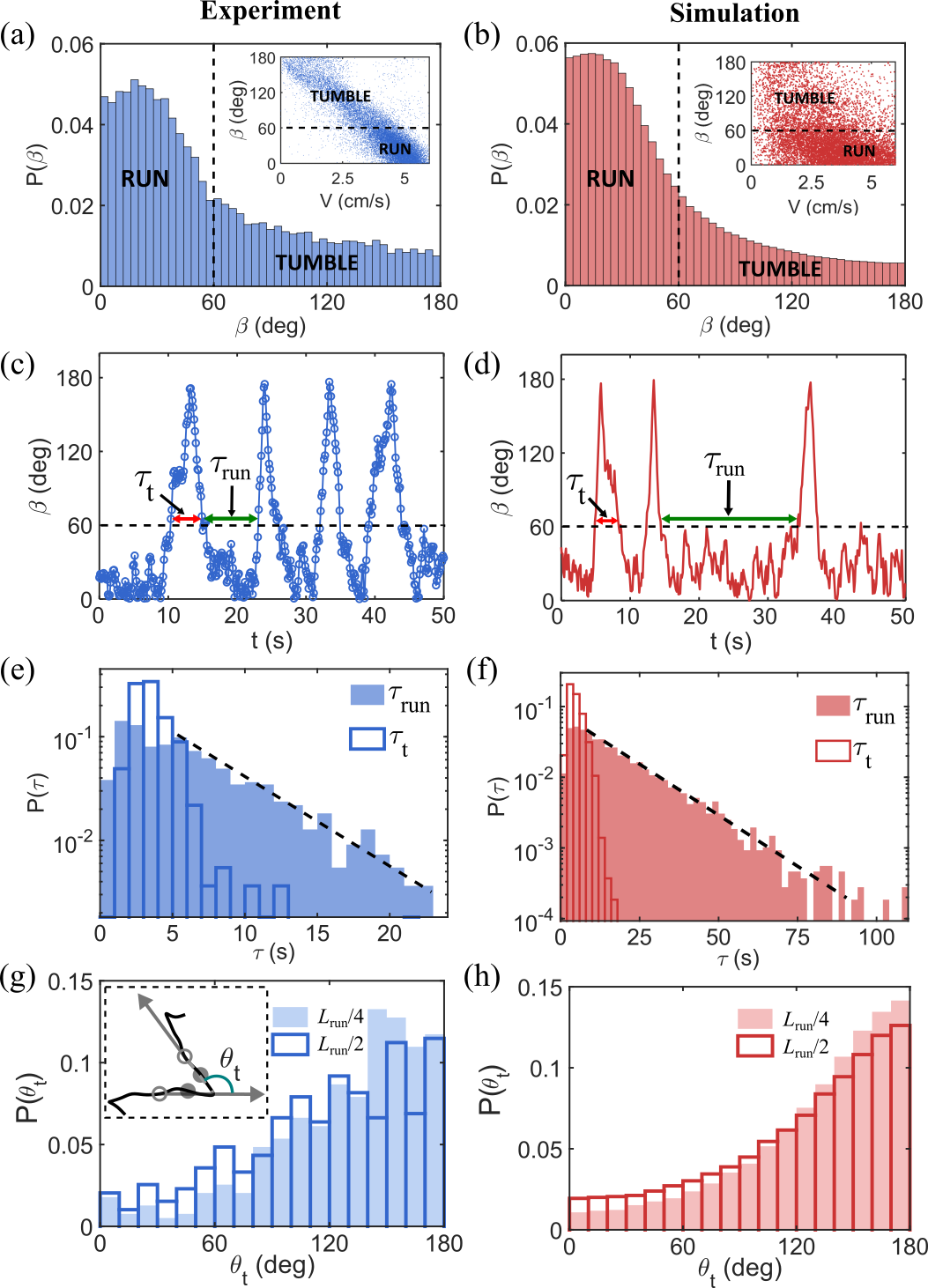}
\caption{\textbf{(a, b)} Probability distribution of $\beta$ shows a pronounced peak near $\beta$ = 0$^\circ$. Insets: $\beta$ and $V$ are inversely related to each other for both experiment and simulation. Black dashed lines at $\beta$ = 60$^\circ$ set a threshold to differentiate between run and tumble events. \textbf{(c, d)} Typical behaviour of $\beta$ as a function of time. $\tau_\mathrm{t}$ and $\tau_{\text{run}}$ represent tumble duration and run-time respectively. \textbf{(e, f)} In both experiment and simulation, we observe exponentially decaying run times (dashed lines as a guide) and unimodal tumble durations. \textbf{(g, h)} Tumble angle ($\theta_\mathrm{t}$) distributions from experiment and simulation respectively. Inset: A schematic defining $\theta_\mathrm{t}$ as the change in orientation between two successive run events. Hollow and solid symbols correspond to measurements taken at half and one-quarter of the run trajectory length (\textit{L}\textsubscript{run}), respectively. Distributions are steeper for $L_\text{run}/4$, indicating sharp run reversal events as predominant tumble events. Experimental and simulation parameters are $v_a$ = 5 cm s$^{-1}$, $D_r$ = 0.06 rad$^2$ s$^{-1}$, $\delta$ = 3 cm, and $\alpha = 90^\circ$.}

	\label{fig2}
\end{figure}
%-----------------------------------------------------%%

To carry out statistical analysis of the RT motion, we record long-time trajectories comprising hundreds of running and tumbling events for experiments and simulations. In experiments, we eliminate trajectories that experience disruptions due to boundary walls (see SM section IC for more details \cite{supp_mat}). We find that two key parameters effectively describe the RT motion: the pair angle $\beta (t)$ representing the instantaneous difference in the in-plane orientations of two robots [see schematic Fig.~\ref{fig1}(c)] and the instantaneous speed $V(t)$ of the centroid $C$. A typical distribution of $\beta$ is shown in Figs.~\ref{fig2}(a) and~\ref{fig2}(b) for experiment and simulation, respectively. We also find that these parameters are inversely related to each other [insets of Figs.~\ref{fig2}(a) and~\ref{fig2}(b)] with a cluster of points at high $V$ and low $\beta$, which we identify as run states and vice versa for tumbles. To differentiate runs from tumbles quantitatively, we set an arbitrary threshold of $\beta = 60 ^\circ$, corresponding to the half-maximum of the run peak. The typical behaviour of $\beta$ as a function of time is shown in Figs.~\ref{fig2}(c) and~\ref{fig2}(d) for experiment and simulation, respectively. By using $\beta = 60^\circ$ as a reference line, we quantify tumble duration ($\tau_\mathrm{t}$) and run-time ($\tau_\text{run}$) from this time series for both experiment and simulation. The resulting distributions $\tau_\text{run}$ show exponential decay in each case [Figs.~\ref{fig2}(e) and~\ref{fig2}(f)]. See SM Fig.~S4 for zoomed-in $\tau_\mathrm{t}$ distributions \cite{supp_mat}. Encouragingly, such exponential distributions in $\tau_\text{run}$ are ubiquitous in many microorganisms like swimming bacteria \cite{berg1972chemotaxis}, algae \cite{polin2009chlamydomonas}, and amoeba \cite{li2008persistent}.  Our experiments also reveal that runs last significantly longer than tumbles [$\left\langle\tau_\text{run}\right\rangle = 7$ s, $\left\langle\tau_\mathrm{t}\right\rangle = 3.5$ s for data presented in Fig.~\ref{fig2}(c) and SM section IE and Fig. S5 for other values of $D_r$, $\alpha$ $\&$ $\delta$ \cite{supp_mat}], consistent with observations made in real biological systems \cite{berg1972chemotaxis, polin2009chlamydomonas}. We also quantify tumble angles, defined as the angle subtended between two successive run directions ($\theta_\mathrm{t}$, see inset to Fig.~\ref{fig2}(g)). Since the run trajectories are not perfectly straight, we use $\theta_\mathrm{t}$ as the difference in orientation angles measured at the midpoints of two successive run events ($L_\text{run}$/2, hollow circles). By analyzing hundreds of tumble events (see SM Fig.~S6 \cite{supp_mat}), we find that the majority of tumbles are sharp, direction-reversing turns, closely resembling those observed in experiments on \textit{Chlamydomonas} \cite{polin2009chlamydomonas}. This is quantified in Fig.~\ref{fig2}(g) and~\ref{fig2}(h) for experiment and simulation respectively, showing tumbles by $180^\circ$ being the most likely events. To further support this conclusion, we modify the run-length to $L_\text{run}/4$ to define $\theta_\mathrm{t}$ (solid symbols in Fig.~\ref{fig2}(g) inset). This allows us to zoom in closer on tumble events, causing noticeably steeper distribution, with a significantly higher frequency of reversals.

%%-----------------FIGURE III---------------%%
\begin{figure}[t]
\centering
	\includegraphics[width=.485 \textwidth]{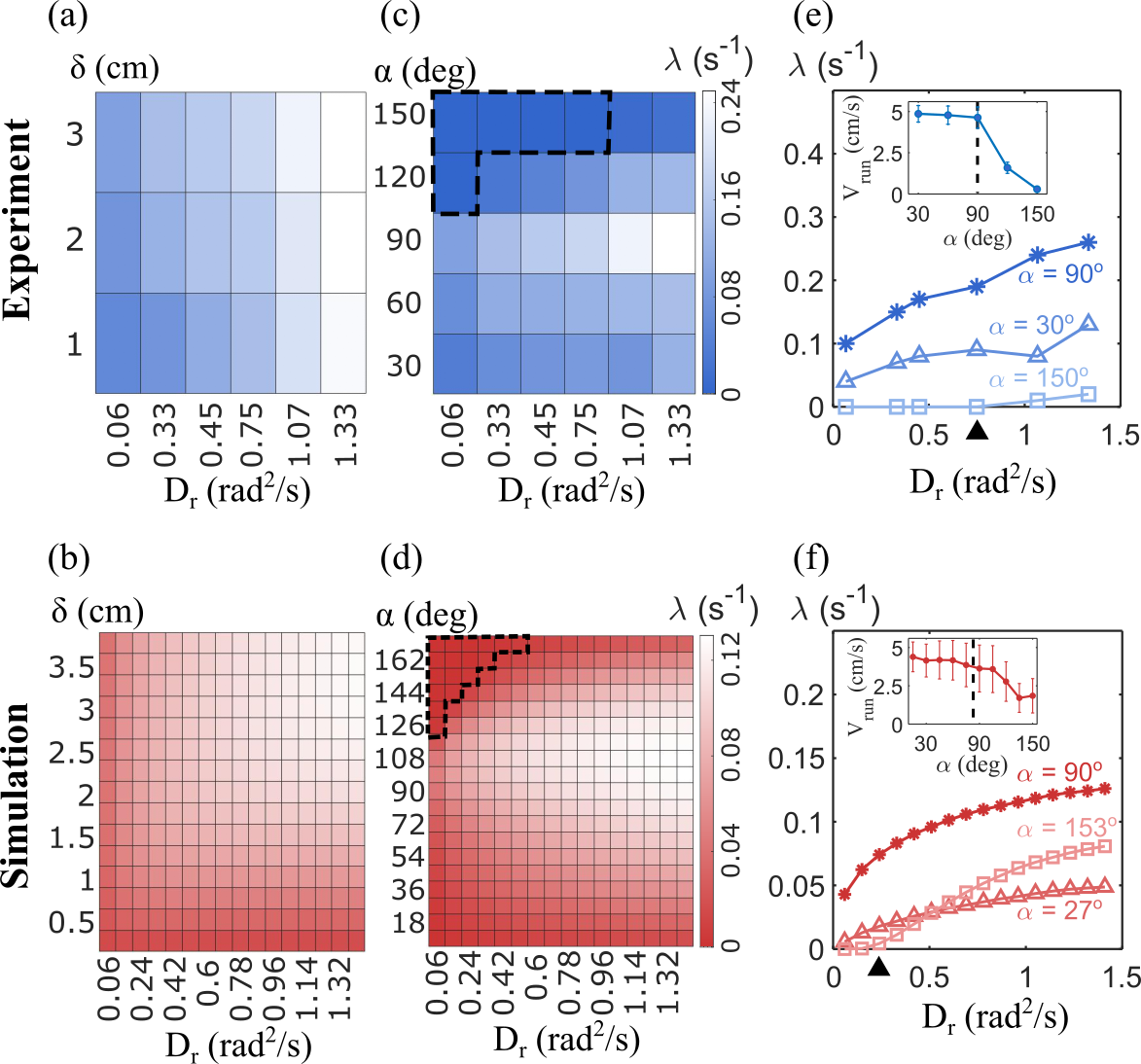}
\caption{\textbf{(a, b)} The tumbling frequency, $\lambda$ increases with $D_r$ and $\delta$. \textbf{(c, d)} The phase diagram of $\lambda$ in $\alpha-D_r$ plane for $\delta$ = 3 cm. The region enclosed by the dashed line corresponds to $\lambda \approx 0$. \textbf{(e, f)} For $\alpha > 90^\circ$, $\lambda$ exhibits a critical $D_r$ indicated by a black arrowhead beyond which tumbling emerges in the system. Insets: The run speed, $V_\text{run}\approx v_a$ for $\alpha \leq 90^\circ$ but decreases monotonically for $> 90^\circ$. Here $D_r$ = 0.06 rad$^2$ s$^{-1}$ and error bars represent standard deviation.}

	\label{fig3}
\end{figure}
%-----------------------------------------------------%%
Inspired by theoretical frameworks used for modeling RT motion ~\cite{cates2008statistical,cates2010prl,cates2013active,cates2014chemotactic,cates2015active,cates2015motility}, we quantify this motion in terms of tumbling frequency, denoted by $\lambda$, which equals $<\tau_\text{run}^{-1}>$. The averaging is performed over all run events. We find that for $\alpha = 90 ^\circ$, $\lambda$ shows a systematic variation as a function of $\delta$ and $D_r$, as shown in Fig.~\ref{fig3}(a) and~\ref{fig3}(b) for experiment and simulation, respectively. Our results show that both $D_r$ and $\delta$ promote tumbling in the system along the expected lines as they introduce stochasticity in the run state and increased torque, respectively \hs{[see Eq.~(14)b of SM \cite{supp_mat}]}. The system also shows an interesting phase behaviour when we vary the parameter $\alpha$. Experiments and simulation show that the run becomes considerably stable for $\alpha > 90^\circ$. This is indicated by regions surrounded by dashed lines corresponding to vanishing $\lambda$ in Figs.~\ref{fig3}(c) and (d) for experiment and simulation, respectively. When we plot $\lambda$ as a function of $D_r$ for various values of $\alpha$, we find a well-defined critical $D_r$ beyond which $\lambda$ increases abruptly, indicated by black arrowheads in Figs.~\ref{fig3}(e) and~\ref{fig3}(f). On the contrary, for $\alpha \leq 90^\circ$, $\lambda$ shows a gradual increase with $D_r$. We further observe that the mean run speed, $V_{\text{run}}$, calculated by taking the average of $V$ while running, remains independent of $\alpha$ for $\alpha\leq 90^\circ$ and decreases with $\alpha$ as $\alpha$ exceeds $90^\circ$ [insets of Fig.~\ref{fig3}(e) and (f)]. 

Next, we perform a theoretical analysis to explain how run and tumble states emerge in our system, thus providing a rationale for the $\alpha - D_r$ phase diagram. The analysis is carried out in terms of the generalized coordinates $\theta_\pm = (\theta_1 \pm \theta_2)/2$, where $\theta_1$ and $\theta_2$ denote the robots’ orientations relative to $\mathbf{N}$ [see Fig.\ref{flowprof}(a)], with the complete treatment provided in the end matter. Our analysis highlights subtle differences in the run states as the parameter $\alpha$ is varied. More specifically, we find that in the plane of $\theta_+$ $\&$ $\theta_-$, runs correspond to fixed stable points, semi-stable, and a stable fixed line for $\alpha > 90^\circ$, $\alpha = 90^\circ$, and $\alpha < 90^\circ$, respectively. As a result, runs correspond to stable configurations requiring a finite $D_r$ value to transition into a tumbling state in accordance with the observation of $\lambda \approx 0$ regions in Figs.~\ref{fig3}(c) and \ref{fig3}(e). Moreover, the geometry of these stable configurations also predicts the run speed scales as $v_a \sin \alpha$, offering an explanation for the observed decrease in run speed as $\alpha$ exceeds $90^\circ$ [see inset plots in Figs.~\ref{fig3}(e) and \ref{fig3}(f)].

Finally, in \textit{Chalmydomonas}, the distal fiber connecting basal bodies actively regulates orientations of beating flagella that influence its swimming behaviour \cite{hayashi1998real}. Interestingly, in our system, parameters $\delta$ and $\alpha$ can be viewed as playing a similar role by altering the orientation of self-propelling robots relative to the connecting rod. Leveraging this, we simulate a scenario (SM movie 6 \cite{supp_mat}) where our system can transition between RT regimes with significantly different tumbling rates, demonstrating possible tuning strategies employed by the real organism. Thus, our mechanical robotic system not only reproduces the key statistical features of RT motion, but it also accounts for its tunability in real living organisms.

%-----------------------------------------------------------------
%       CONCLUSION
%-----------------------------------------------------------------
To conclude, inspired by microorganism motility, we present a robotic model system that replicates run-and-tumble-like (RT-like) motion. Our robots, driven by motorized wheels through mechanical gears and responding to microcontroller signals instantaneously, undergo a rolling-without-slipping motion. As a result, they faithfully emulate overdamped active motion, a hallmark of microorganisms operating at extremely low Reynolds numbers. We program both robots to exhibit an overdamped active Brownian (AB) motion. The rigid rod, attached to a pivot point on the robot's body, rotates freely and facilitates rotational motion. To capture the role of the contractile fiber connection between flagella in the \textit{Chlamydomonas} microswimmer, we vary the pivot placement relative to the robot's polarity axis, leading to a wide range of tunable RT-like dynamics. We quantify this RT-like motion using the tumbling frequency ($\lambda$) and identify two key tuning parameters. The first, $\delta$, is the distance of the pivot from the robot's centre, which increases $\lambda$ by amplifying torque. The second, $\alpha$, is the angle between the pivot-to-center line and the polarity axis, which also affects the $\lambda$ depending on rotational noise programmed inside each robot. We also developed a theoretical model that captures the essential features of our system and reproduces the experimental results, elucidating the intricate dependence of RT dynamics on $\delta$ and $\alpha$. 

\textit{Acknowledgements} - NK acknowledges financial support from DST-SERB for CRG grant number CRG/2020/002925 and IITB for the seed grant. HS acknowledges SERB for the SRG (grant no. SRG/2022/000061-G). SP thanks CSIR India for the research fellowship (File no: 09/087(1040)/2020-EMR-I). \\

\textit{Data availability} - 
The data that support the findings of this Letter are openly available at \cite{data}. The original data exists in the form of experimental videos, which will be available from the corresponding authors upon reasonable request.
%%-------APPENDIX-------------%%
% \appendix

%% REFERENCES
\bibliography{references}% Produces the bibliography via BibTeX.

\appendix
\section{End Matter}

\begin{figure*}
	\includegraphics[width= .8\textwidth]{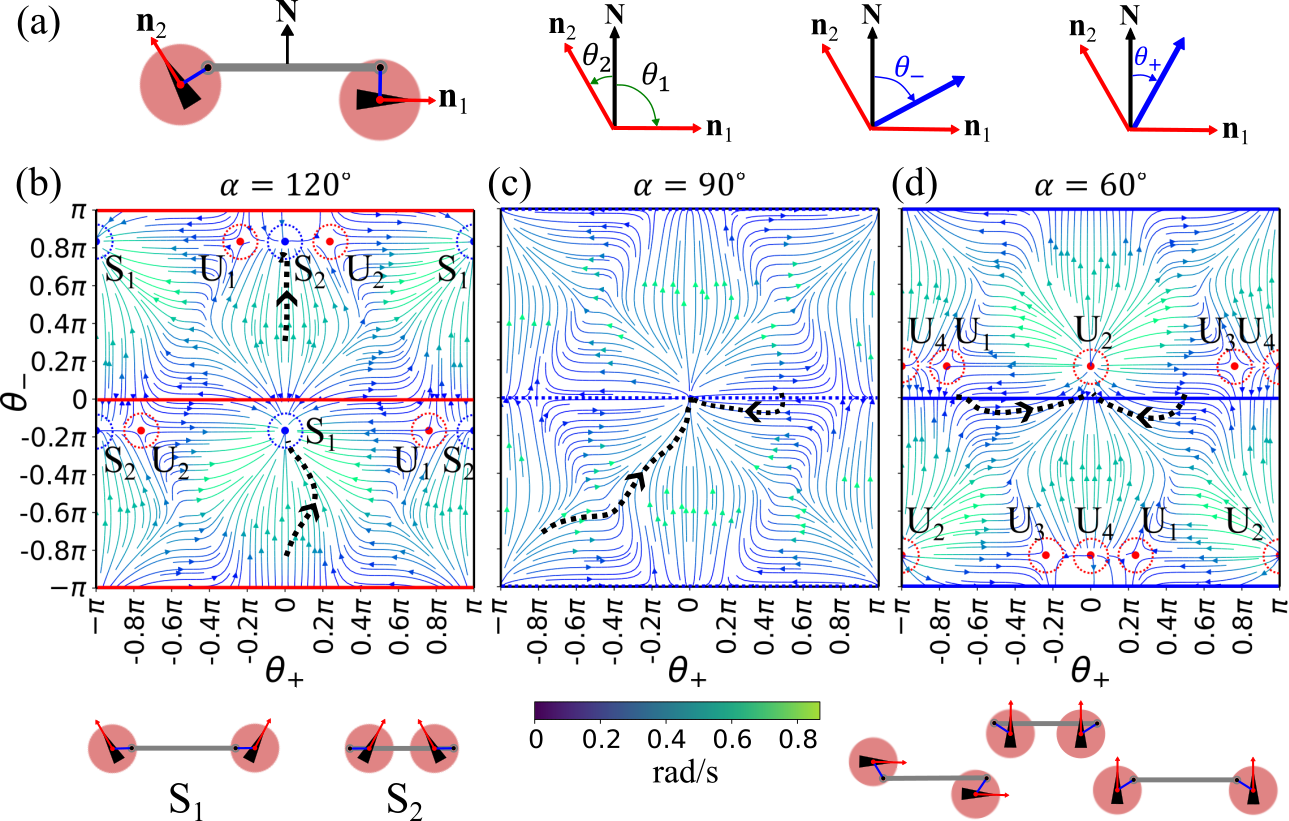}
\caption{\textbf{(a)} A schematic diagram clarifying the angular coordinates $(\theta_+,\theta_-)$. Here, $\theta_\pm = (\theta_1\pm\theta_2)/2$.  \textbf{(b)} Flow diagram of  $(\theta_+,\theta_-)$ for $\alpha = 120^\circ$ evaluated from the deterministic part $\mathcal{T}_\pm$ of Eq. ~\eqref{theq}. $S_1$ $\&$ $S_2$  and $U_1$ $\&$ $U_2$ encircled by dotted lines represent stable and unstable points, respectively. Black dashed lines represent experimental trajectories for $D_r = 0$. Schematic configurations of $S_1$ and $S_2$ are shown at the bottom. Here, $\mathbf{n_1}$ and $\mathbf{n_2}$ are misaligned by an angle of $2\alpha-\pi=60^\circ$ in both cases. \textbf{(c)} Flow diagram for $\alpha = 90^\circ$. Blue dotted lines at $\theta_- = n\pi$ are semi-stable in $\theta_-$ and neutral in $\theta_+$, where $n$ denotes positive integers.  Black dashed lines are the experimental trajectories for $D_r = 0$. %The system predominantly runs along the dotted line where $\mathbf{n_1}\parallel \mathbf{n_2}$.
 \textbf{(d)} Flow profile for $\alpha = 60^\circ$. The system exhibits four unstable points ($U_1$, $U_2$, $U_3$, and $U_4$). Solid blue lines at $\theta_- = n\pi$ represent configurations that are stable in $\theta_-$ and neutral in $\theta_+$. The black dashed experimental trajectories for the $D_r = 0$ case. Three among infinitely many run configurations corresponding to $\mathbf{n_1}\parallel \mathbf{n_2}$ are shown at the bottom. The color bars in (b), (c) $\&$ (d) correspond to the magnitude of the vector $(\mathcal{T}_+, \mathcal{T}_-)$ in rad s$^{-1}$.}
  
	\label{flowprof}
\end{figure*}

Here, we perform a theoretical analysis to explain the observed RT dynamics in our system and to provide a rationale for the $\alpha-D_r$ phase diagram. As observed in experiments and numerical simulations, the dynamics of the system at any given time are entirely governed by its internal configuration, which is represented by variables $\theta_i$ ($i$ = 1 and 2 for two robots) [Fig.~\ref{flowprof}(a)]. Therefore, we define another set of generalized angular coordinates: $\theta_\pm = (\theta_1 \pm \theta_2) / 2$ (see Fig.~\ref{flowprof}(a)). Note that $\beta$ equals the principal value of the angle $2\theta_-$ in the range $[-\pi, \pi]$ and $\theta_+$ indicates the average orientation of the robots with respect to the rod. Fig.~\ref{flowprof}(a) clearly shows that motility causes increasing and decreasing trends in $\theta_1$ and $\theta_2$, respectively. Therefore, by definition, $\theta_-$ also increases with time over large time scales. Moreover, for integers $n$, when $\theta_- =  n\pi$, the robots are parallel to each other and thus move with the maximum possible speed of $v_a$. Similarly, when $\theta_- = (2n+1)\pi/2$, the system shows tumbling behavior. To elaborate further, we derive the equations of motion for $\theta_{\pm}$ from \hs{Eqs.~(30) and (32) of SM \cite{supp_mat}}, which are given as follows (also see SM section IID \cite{supp_mat}):
\begin{equation}\label{theq}
    \dfrac{d\theta_\pm}{dt}=\mathcal{T}_\pm+\sqrt{2D_\pm}\eta_\pm(t),\\
\end{equation}
\hs{where $\mathcal{T}_+=(2v_a/\ell)\sin\theta_-\left(\sin\theta_+ -\ell\cos\theta_+\mathcal{G}\mathcal{H}\right)$, $\mathcal{T}_-=-(2\delta v_a/\Gamma_\tau)\mathcal{G}\cos^2\theta_+\cos(\theta_-+\alpha)\sin\theta_-$,}
%\begin{eqnarray}\label{tp}
% \mathcal{T}_+&=&2v_a\sin\theta_-\left(\dfrac{1}{\ell}\sin\theta_+ -\cos\theta_+\mathcal{G}\mathcal{H}\right)\\\label{tm}
% \mathcal{T}_-&=&-\dfrac{2\delta v_a}{\Gamma_\tau}\mathcal{G}\cos^2\theta_+\cos(\theta_-+\alpha)\sin\theta_-
%\end{eqnarray}
and $D_\pm$,  $\mathcal{G}$ and $\mathcal{H}$ are the functions $\theta_+$ , $\theta_-$, and $\alpha$ (see SM section IID \cite{supp_mat}).  The strengths of the noise terms $D_\pm$ in Eq.~\eqref{theq} are of the order of $D_r$. \hs{$\eta_\pm(t)$ are the delta-correlated noise with zero mean and variance one.}

%%-----------------FIGURE IV---------------%%
%-----------------------------------------------------%%

Let us now analyze the effect of the deterministic components, $\mathcal{T}_\pm$, on the dynamics of $\theta_\pm$.  In Figs.~\ref{flowprof}(b)-\ref{flowprof}(d), we display the flow profiles of the angular coordinates $(\theta_+, \theta_-)$ in the absence of noise. The color map represents the magnitude $\mathcal{T}$ in rad s$^{-1}$ of the vector with two components $(\mathcal{T}_+, \mathcal{T}_-)$. Note that the system is invariant under the transformations $\theta_1 \to \theta_1 \pm 2\pi$ and $\theta_2 \to \theta_2 \pm 2\pi$. Consequently, it remains invariant under $(\theta_+, \theta_-)\to(\theta_+ \pm \pi, \theta_- \pm \pi)$ and $(\theta_+, \theta_-)\to(\theta_+ \pm \pi, \theta_- \mp \pi)$. This implies that the first and third quadrants, as well as the second and fourth, represent the same set of systems. This symmetry is evident in the flow profiles as well.

We begin with the case where $\alpha > 90^\circ$, with a typical flow profile shown for $\alpha=120^\circ$ in Fig.~\ref{flowprof}(b). Here, we find two stable ($S_1$, $S_2$) and two unstable ($U_1$, $U_2$) fixed points. We also observe an unstable fixed line along $\theta_- = n\pi$ which is less relevant to the results presented. Locations of stable points in the first and second quadrants are $(\theta_+, \theta_-)$ = $(\pi, 3\pi/2 - \alpha)$ and $(0, 3\pi/2 - \alpha)$ for $S_1$ and $S_2$ respectively, with system configurations shown below Fig.~\ref{flowprof}(b). Both configurations correspond to run states. Therefore, the system favors these stable points for low noise values, resulting in a pronounced run regime at low $D_r$ values, explaining the experimental results presented in Figs.~\ref{fig3}(c) and (e).  Interestingly, both $S_1$ $\&$ $S_2$ correspond to non-aligned robots ($\mathbf{n_1} \nparallel \mathbf{n_2}$). %Interestingly, in this case, the stable run states do not correspond to aligned robots (as was the case for $\alpha = 90^\circ$), but rather to anti-aligned vectors $\mathbf{s}_i$. 
To test this, we conduct experiments for the $D_r = 0$ case using random initial conditions in $(\theta_+, \theta_-)$ and find that the system follows the predicted flow lines (black dashed lines in Fig.~\ref{flowprof}(b)) before eventually settling into one of the stable regions, indicated by the dashed circular regions (see SM movie 7 \cite{supp_mat}). Moreover, as evident from configurations $S_1 \& S_2$, the system's run speed at the stable points is given by $v_a \sin \alpha$, which decreases with $\alpha$ for $\alpha > 90^\circ$. This explains the decreasing trend in $V_\text{run}$ shown in the insets of Figs.~\ref{fig3}(e) and~\ref{fig3}(f). We also observe that at low but finite $D_r$ values, the system runs corresponding to $\theta_-$ = $\pi/2 - \alpha + n\pi$, which equals $\theta_- =n\pi - \pi/6$ for $\alpha = 120^\circ$ while occasionally transitioning between $S_1$ and $S_2$ leading to tumbling events [See SM Section IIE \cite{supp_mat}].

%escapes its stable points, causing it to jump between stable states by a value of $\pi$ leading to a tumbling event. 
 
%An alternate way to visualize RT dynamics is through the time dependence of the parameter $\theta_-$. In Fig.~\ref{flowprof}(e), which corresponds to $D_r$ = 0.06 rad$^2$ s$^{-1}$ in simulation, we find that $\theta_-$ always plateaus around $\pi/2 - \alpha + n\pi$, which equals $\theta_- =n\pi - \pi/6$ for $\alpha = 120^\circ$ shown by gray dotted lines. However, due to rotational noise, the system occasionally escapes its stable points, causing it to jump between stable states by a value of $\pi$ leading to a tumbling event. As $D_r$ increases, the system is more likely to escape the stable points sooner, leading to an increased tumbling rate, as observed in both simulations and experiments. 

For $\alpha = 90^\circ$, no stable fixed points are observed. Instead, there exists a fixed line [blue dashed line in Fig.~\ref{flowprof}(c)] at $\theta_- = n\pi$, which is semi-stable in $\theta_-$ and neutral in $\theta_+$. Therefore, as the system reaches this line, it runs for a while before sliding towards increasing $\theta_-$ due to noise. Experiments performed at $D_r = 0$, in this case, show traced paths that agree well with theoretical values (see SM movie 8 \cite{supp_mat}). %Also, as shown in Fig.~\ref{flowprof}(f), the system still tends to stay closer to $\theta_- = n\pi$ (dotted gray lines) due to the neutral nature, although only for a short period, describing the origin of RT dynamics on our system for $\alpha = 90^\circ$ shown in Fig.~\ref{fig1}. 
Finally, for $\alpha < 90^\circ$, and using $\alpha = 60^\circ$ as a typical example, we find four unstable points ($U_1$, $U_2$, $U_3$, $U_4$) and a fixed line at  $\theta_- = n\pi$ as shown in Fig.~\ref{flowprof}(d). This fixed line is neutral in $\theta_+$ while stable in $\theta_-$, corresponding to infinitely possible configurations with $\mathbf{n_1} \parallel \mathbf{n_2}$, with schematics of three typical configurations shown below in Fig.~\ref{flowprof}(d). Consequently, $V_\text{run}\approx v_a$ for $\alpha \leq 90^\circ$ as observed experimentally in inset of Fig.~\ref{fig3}(e). Again, the grey dashed line indicates the experimental trajectory for $D_r = 0$ (SM movie 9 \cite{supp_mat}). %Furthermore, in Fig.~\ref{flowprof}(g), we present $\theta_-$ as a function of time, which exhibits relatively noisier plateaus corresponding to run states with occasional escapes, equivalent to the behavior observed for $\alpha > 90^\circ$. A key distinction, however, is that while the system for $\alpha > 90^\circ$ is limited to two stable run states, an infinite number of run configurations are possible when $\alpha < 90^\circ$.
%Therefore, our theoretical model not only explains the emergence of complex RT motion in experiments but also elucidates how this complex behavior systematically depends on experimental parameters $\alpha$ and $D_r$.

\end{document}